# Non-twisted stacks of coated conductors for DC magnets: analysis of inductance and AC losses

*Davide Uglietti[a]\*, Rui Kang[a,b], Rainer Wesche[a], Francesco Grilli[c]*
[a] *Ecole Polytechnique Fédérale de Lausanne (EPFL), Swiss Plasma Center (SPC), CH-5232 Villigen PSI, Switzerland*
[b] *Department of Engineering and Applied Physics, University of Science and Technology of China, Hefei 230026, China*
[c] *Karlsruhe Institute of Technology, Karlsruhe, Germany*
\* Corresponding author. E-mail address: davide.uglietti@psi.ch

**Abstract** - In the last 10-15 years, the most common strategy in the development of High Temperature Superconducting (HTS) cable for magnets has been to imitate Low Temperature Superconducting (LTS) cable designs. However, requirements for LTS materials are not valid for HTS materials, which are extremely stable. For example, non-twisted multifilamentary Bi-2223 tapes have been successfully used in several magnets. This paper review stability and analyse inductance and AC losses in non-twisted stack of HTS tapes. Numerical calculations show that twisting has negligible effect on inductance variations in a stack of tapes. Regarding AC losses, any magnet built with coated conductors have larger losses than LTS ones, because of the aspect ratio and large width of the tape. If a wide tape is replaced by a non-twisted stack of narrow tapes, losses and residual magnetisation could be reduced. In contrast with multifilamentary wires, twisting a stack of tapes reduces losses only marginally. Therefore, cables composed of non-twisted stack could be designed to have losses comparable to the one of twisted stack concepts. Few examples of large cables for fusion applications are discussed. Designs based on non-twisted stacks can be simpler, more robust and cost effective than twisted ones.

## 1. Introduction

Cable and conductors made of RE-Ba$_2$Cu$_3$O$_{7-\delta}$ (REBCO) tapes have attracted more and more interest in the last 10-15 years, see reviews [1], [2] and [3]. Most of the proposed HTS designs imitate the twisted/transposed concepts that were developed for LTS cables in the 70's and 80's (see [4] for a review of LTS fusion cables and conductors). Therefore, it is important to discuss the reasons that guided the design of LTS cables and whether their requirements should be applied also to HTS materials.

In the 50's and 60's, the main obstacle to the construction of superconducting magnet was the low stability of LTS materials. The solution was to subdivide the superconducting material in very fine twisted filaments, embedded in a low resistance matrix. When large currents were needed, it was found that simply winding several strands in parallel would lead again to instability, because the strands would be fully coupled, behaving like a large monolithic conductor. An example is the conductor of the T-7 magnet [5], which was composed of several parallel NbTi strands. This magnet reached only 90% of the expected performance, due to flux jumps and large saturated losses. To avoid that, twisted/transposed designs have been introduced during the 70's and 80's.

There are few exceptions to the rule of twisted/transposed conductor: 1) The LIN-5 conductor [6], which is composed of 25 parallel NbZr wires arranged in a ribbon-like conductor. 2) The IMP mirror coils [7], which were wound with a 2.9 mm x 1.4 mm conductor composed of 15 NbTi parallel strands in copper matrix. For both magnets, no operating anomalies have been reported. 3) In the last decades, many magnets have been built with Bi$_2$Sr$_2$Ca$_2$Cu$_3$O$_x$ (Bi-2223) tapes and have been successfully operated; as well-known to the reader, filaments in high current density Bi-2223 tapes are parallel (non-twisted). If this configuration was applied to LTS tape, it would likely lead to flux jump and saturated loss instabilities, like in the T-7 conductor. These few examples challenge the obligation to twist and/or transpose elements in superconducting conductors, and trigger few questions: if Bi-2223 tapes (non-twisted filaments) can be used in magnets, why a non-twisted stack of two or three coated conductor tapes can not be used? If those non-twisted conductors worked because of the small size, what is the maximum size for non-twisted conductors?

This paper tries to examine the reasons for twisted and transposed HTS cable. The main objections against non-twisted conductors are stability, variation in inductance (leading to current unbalance) and AC losses. In section 2, stability in LTS and HTS magnets is reviewed. In section 3 the inductance variation in the so-called "partially transposed" cables, like twisted and non-twisted stacks of coated conductors, is studied. In section 4 the ramping losses for small solenoid built with single tape and with non-twisted stacks are analysed and discussed. In section 5 the losses in large cables (twisted and non-twisted stacks of tapes) are compared, considering the application in Fusion magnets.

## 2. Stability in LTS and HTS conductors and magnets – a brief review

LTS materials have low stability against thermal disturbances. The reason is the very fast decrease of the critical current density, $J_c$, with temperature and the very steep superconducting transition with temperature. The stability issue was solved with the subdivision of the superconducting material in fine, twisted filaments, which should be embedded in a highly conductive matrix (for example high purity copper); twisting of the filaments is required to uncouple them magnetically.

In contrast with LTS, HTS have a larger temperature margin, $J_c$ decreases very slowly with temperature and the transition is smooth; in addition, the large temperature margin and the fact that specific heat grows with $T^3$ at low temperatures (4.2 K – 10 K) strongly supress the positive feedback loop leading to quench. Therefore, they are much more stable against thermal disturbances. Indeed, the phenomenology of HTS strands and magnets is completely different from the one of the LTS counterparts:

- "Training" is a common phenomenon observed in most of LTS magnets, and it is due to tiny wire movements or epoxy cracking which cause quenches. Instead, there are no reports (to our knowledge) of training or of quenches by wires movement / epoxy cracking in HTS magnets.
- In LTS filamentary wires, the filament diameter has to be less than few tens of micron to guarantee sufficient stability. These values are obtained using equation 7.7 in [8], page 134: $\frac{\mu_0 J_c^2 a^2}{3 C_V (T_c - T)} < 1$, where $J_c$ is the critical current density in the superconductor, $a$ the filament diameter, $C_V$ the volumetric specific heat and $T_c$ the critical temperature. When the same formula is applied to HTS (see for example [9] and [10]), the maximum diameter is several mm, i.e. two to three order of magnitude larger than in LTS. In LTS the fine filaments must be twisted with a twist pitch shorter than the critical twisting length $L_c = \sqrt{\frac{d \rho J_c}{dB/dt}}$ (equation 8.36 in [8], page 175), where $d$ is the filament thickness, $\rho$ the matrix resistivity, $dB/dt$ the magnetic field sweep rate. If the twist pitch is $\geq L_c$, the filaments would be magnetically coupled, and stability and AC loss would be similar to the ones of a single, very large filament.
- Flux jumps in tapes: most of the magnets build in 60's and 70' with $Nb_3Sn$ tapes reached only 20% to 50% of the field expected from short sample $J_c$, the reason was flux jump instabilities [11]. To mitigate that, large cross section of high purity (RRR>1000) Al stabilizer had to be added and, even more important, cooling had to be optimal [12]. The flux jump fields of HTS tapes was studied in [13]: "*With flux jump fields being between one and two orders of magnitude higher than for conventional SC, the necessity for the use of a multi filamentary conductor is greatly relaxed. Therefore HTSC in the form of solid tapes become an attractive option for many applications*".

Flux jumps occur when $B > \sqrt{3\mu_0 C_V (T_c - T_0)}$: at 4.2 K the right term is larger for HTS material because of the higher critical temperature, and it increases even more at high temperatures ($C_V \sim T^3$). Indeed, there is no report, to our knowledge, of flux jump instabilities in coated conductor tapes even at 4.2 K, high field and high current densities.

- In LTS wires a high conductive matrix (for example copper with RRR>100) must be present and should be in good contact with the filaments; moreover, in some LTS magnets, heat removal had to be very good (leading to cable in conduit concepts). Instead in coated conductor, there are no reports of instabilities, even if the Cu layer is very thin and has a poor RRR (<40) and in conduction cooled magnets (cryogen free).

HTS materials are more stable than LTS at 4.2 K; the stability is even increased at temperatures > 20 K. The microstructure of Bi-2223 tapes, $Bi_2Sr_2CaCu_3O_x$ (Bi-2212) wires and REBCO coated conductors is not determined by stability, but by the maximization of the critical current and the manufacturability in long length. For example, Bi-2223 tapes are composed of filaments because the $J_c$ is higher than in a mono-core tape; the reason is the better texturing of the ceramic close to the Ag matrix. It is technically possible to twist the filaments in Bi-2223 tapes (AC loss reduction only at low frequency), and even introduce resistive barriers to decrease losses at any frequency, but twisted multifilamentary tapes had lower current densities and higher cost. Consequently twisted tapes were seldom (probably never) used in magnets and are not produced anymore by Sumitomo.

The countermeasures to wire movement, magnet training, flux jumps, and other instabilities affecting LTS should not be applied "a priori" to HTS, because HTS are intrinsically immune to these issues. Therefore, it would be reasonable to expect that the same criteria that have driven the HTS strands development (maximization of $J_c$) would also be used for HTS cable development. Instead, most of the HTS cable designs proposed so far are imitating LTS cable designs. Probably the only exception is the CORC, which exploit a unique feature of coated conductors (tolerance to large compressive longitudinal strain and asymmetric layout of the tape). All other HTS cable designs are merely copied from the LTS ones. This means that the cures to instability issues, which are specific of LTS materials, have been applied also to HTS, even if not needed.

## 3. Inductance

The terms "transposition" and "twisting" are used frequently when discussing superconducting cables, but they have not the same meaning. "transposed" means that all the strands follow the same trajectory (if strands are translated in longitudinal direction), or, in other words, that the strands periodically swap their positions; it follows that transposed strands have the same inductance. "twisted" means that rotation is applied to the strands during cabling. In Rutherford cables and ITER cables, the strands are twisted and transposed. In Roebel cables (or bars), the strands are non-twisted but transposed. Examples of twisting without transposition are: tapes in CORC, tapes in twisted stacks, strands in the sub-cable of the RW2 conductor for EU-DEMO [14] and even filaments in $Nb_3Sn$ and NbTi wires.

In papers discussing twisted, non-transposed cables and conductor, the term "partially transposed" is often employed. The meaning of "partially transposed" is vague, and is never discussed how "partial transposition" should be quantified, measured or estimated. It is important to recall that the scientific method employed in the last three centuries is based on measurable evidence: if something is not measurable or quantifiable (i.e. can be expressed with numbers), should not find place in science and technology. Quoting Lord Kelvin: "When you can measure what you are speaking about and express it in numbers, you know something about it", seen at Hunterian Museum, University of Glasgow.

How to quantify transposition? In transposed cables, all strands have the same inductance, then the amount of "partial transposition" can be quantified as the variation in inductance among the strands. To our knowledge, only in [15], section 6, the calculation of the inductance variation for a straight, non-twisted stack of tapes has been attempted. The total inductance of tape $i$ in the stack ($N$ tapes) is the sum of the tape self-inductance plus the sum of the mutual inductances between tape $i$ and all other tapes: $L_i = L_{self} + \sum_{j=1, j \neq i}^{N} M_{ij}$.

We have repeated the calculation, using equation (9) from [16], p. 35 for the self-inductance. The mutual inductance is obtained considering the first three terms of equation (3) from [16], page 33: $M_{ij} = 0.002l \left( \log \frac{2l}{GMD_{ij}} - 1 \right)$, where $l$ is the stack length, and $GMD_{ij}$ is the geometric mean distance between tape $i$ and $j$. The geometric mean distance is derived from Fig. 2 of [17]: $GMD_{ij} = \left( 0.41 \frac{D_{ij}}{w} + 0.22 \right) 2w$, where $w$ is the tape width. An alternative expression for $GMD_{ij}$ can be found in [18], page 322-323. The largest variation in inductance (per unit length) is between the central tape (which has the largest inductance) and the outermost tapes in the stack, and it is about 8% for a stack 12 cm long. If the same calculation is repeated for a longer stack, the mismatch get smaller.

In order to include also the effects of twisting and winding, the inductance has been calculated with numerical methods. We have used the M'C module of the Cryosoft package. A stack of 30 tapes (4 mm wide and 0.1 mm thick), wound in one turn of 1.9 m radius, has been considered, and each tape was modelled with iso-parametric bricks. It has been found that the inductance variation is about 4%, irrespective of the twist pitch (from non-twisted down to 20 cm twist pitch). Almost all the variation in inductance comes from the summation of the mutual inductances. Therefore, at least for this case (stack of tapes in a large coil), twisting does not significantly reduce the inductance variation and thus does not homogenise the current distribution.

The first cabling stage of DEMO RW2 conductor [14], which is composed of "partially transposed" strands, has also been analysed. This first stage is composed of 18 strands (1 central copper wire + 6 $Nb_3Sn$ strands + 12 $Nb_3Sn$ strands) and both layers of wires have the same twist pitch of 95 mm. The variation in inductance between the strands in the inner layer and the one in the outer one is about 2%; if the twist pitch is reduced to 60 mm, the inductance variation becomes 4%; for 20 mm twist pitch, the variation in inductance is 16%.

One more conductor that has been studied is the non-twisted, non-transposed NbTi conductor used for the fabrication of the T-7 tokamak [5]. The conductor is composed of 16 large NbTi strands arranged in two rows (see Fig. 2 in [4]). The maximum variation in inductance between the strands is about 10% (winding radius is 0.5 m). All these values are summarized in Table 1.

To assess whether a value of inductance mismatch is acceptable or not, one should also consider the transverse resistance between the strands, the ramp rate and the temperature margin: the idea is that current redistribution takes place through resistive materials, thus generating heat (see [19]). This analysis should be carried out on a case-by-case basis. Nonetheless, it can be concluded that inductances mismatches of few % are tolerable in most of applications of HTS and LTS cables, because some of the conductors in Table 1 have shown no major issue related to inductances. In addition, HTS cables could probably tolerate larger inductance variations (and thus current inhomogeneity) than LTS cables because they are much more stable, see for example [20]. If the strands are insulated (i.e. exchange current only at terminals), than the inductance mismatch must be much smaller, as reviewed in [19].

Table 1. Estimated inductance variation in non-transposed cables

| cable | Twist pitch | Max inductance variation (per meter) |
|---|---|---|
| **stack of 30 tapes** (wound on R= 1.9 m) | Infinite | 4% |
| | 1000 mm | 4% |
| | 200 mm | 4% |
| **1+6+12 Nb$_3$Sn** (wound on R= 1.9 m) | 95 mm | 2% |
| | 60 mm | 4% |
| | 20 mm | 16% |
| **T-7** (wound on R= 0.5 m) | Infinite | 10% |

## 4. AC loss in small solenoids and dipoles

Before dealing with non-twisted stacks of coated conductors (section 4.2), it is worth to review and discuss the AC losses in coils built with REBCO coated conductors and with commercial Bi-2223 multifilamentary tapes, which are composed of non-twisted filaments.

### 4.1 Hysteretic loss in solenoids

Even if a large number of coils and magnets have been built with coated conductors in the last decade, there are few publications dealing with analysis and/or measurement of ramping loss. A relatively well-studied case is the cryogen-free 25 T magnet at Tohoku University: the magnet consists of a LTS outsert and a HTS insert; see [21] for the characteristics of the coils. This section reviews the results published in [22] and [23], adding minor observations.

In the HTS insert the main loss component is the hysteretic one. The hysteretic loss can be estimated [22-23] as the sum of the loss associated to the axial field component (parallel to the broad face of the tape) plus the loss associated to the radial field component (perpendicular to the broad face of the tape). The infinite slab model is used for both components. This model is applied to each tape for the axial component and to the whole pancake for the radial component, because of the magnetic interaction between tapes. In fact the tapes in the pancake interior (or in a stack) are screened from the outermost tapes (see [22-23]).

In [22-23] the instantaneous loss was calculated and formula (6) in [22] is simply the time derivative of the well-known formula for an infinite slab, see also [8], page 162-163. An additional term accounts for the loss due to transport current (see also [24], page 10-11), but this term is less than 1% for the axial component and <10% for the radial component; therefore it could be neglected for rough estimations.

It was found (see Fig. 6b in [22]) that at the beginning of the field ramp (<5 T) the largest loss contribution originates from the axial field. The reason is that the penetration field is only few tenths of Tesla for the tape oriented parallel to the field. Later during the field ramp, at larger fields (>10 T), the largest contribution to the instantaneous loss comes from the coil volume exposed to large radial components of the magnetic field. The reason is that for pancakes at coil ends the penetration field is several Tesla, and the loss (J/m) keep growing with the square of field (formula (6) in [22]) up to the penetration field.

In [25], the total peak instantaneous loss in the HTS insert was estimated at 6.5 W and at 2.6 W in the LTS. The HTS coil volume can be evaluated from the coil dimensions [21] to about 0.021 m$^3$, while the LTS coil volume is about 0.20 m$^3$. Then the peak instantaneous losses per unit volume are about 315 W/m$^3$ for the HTS insert and 13 W/m$^3$ for the LTS outsert. It should be stressed that these are average values over the whole volume of the coil; the loss density at the HTS coil ends is much larger; for example, the peak instantaneous loss is about 1800 W/m$^3$ (0.5 W from Fig.5 in [23] divided by the volume of one pancake) in the outermost pancakes. In terms of energy density deposited during the ramp, the loss in the whole HTS coil is about 850 kJ/m$^3$. In the top pancake it would reach over 2700 kJ/m$^3$; this value is comparable to the ones of twisted and non-twisted stacks of REBCO tape, as will be shown in section 5.

One more striking difference between the LTS and the HTS coils, is that in the LTS the instantaneous power loss is constant during the whole ramp, because its main contribution is from the coupling losses (proportional to the ramp rate, which is constant). Instead, in the HTS coil, the instantaneous power loss is hysteretic, and grows continuously (power law) during the ramp; the maximum value is reached at the end of the ramp, when the field is the highest and the critical current and temperature margin are the lowest.

The analysis of losses is in agreement with the measurements of the temperature of the coils, as shown in [2] and [25]. The temperature of the HTS coil raised from 4.3 K to about 7.6 K during the energization of the magnet; instead the temperature of the LTS coils increased only to 5.1 K. Obviously the HTS coil could be operated at such high temperature because the HTS magnet have a much larger temperature margin than any LTS magnets. It is important to stress that the cooling capacity does not need to be increased to face the larger heat load. In fact, as discussed in [25], the cooling efficiency of GM cryocoolers rises with temperature: when the operating temperature is raised from 4.2 K to 8 K the cooling capacity increase from 1.5 W to 10 W, of course for the same electric power consumption at room temperature. The modification to the cooling system simply consisted in using two different cooling circuits for the LTS and HTS magnets, so that the two coils can be operate at different temperatures.

A similar analysis was published [26] by the Chinese Academy of Sciences, in Beijing. The magnet is a split coil wound with Bi-2223 tapes. In the Bi-2223 split coil the energy loss per unit volume is about 750 kJ/m$^3$.

### 4.2 Loss reduction

Ramping loss in HTS tape coils is 1 to 2 order of magnitudes larger than in LTS multifilamentary coils, but the enormous amount of heat does not prevent the operation. A more important effect, at least for some applications, is the perturbation of the central field caused by the magnetisation. Therefore, one may wonder if it would be possible to reduce loss (and thus magnetisation). Introducing fine twisted filaments is the usual method for all LTS magnets in the last 40 years. On the other hand, it is not obvious to do that in coated conductors and maintaining large engineering current density at the same time. For example, twisted multifilamentary Bi-2223 tapes had low current density and higher cost than non-twisted ones, hence were abandoned.

There is another strategy that can be used with materials that can not be twisted easily, as discussed by Campbell [24]: "*using a thin conductor … in the direction perpendicular to the field*". In fact, the hysteretic loss is proportional to the width of the conductor, of course at fields larger than the penetration field; the width is defined as the projection of the conductor in the plane perpendicular to the field direction. It follows that by modifying the aspect ratio of the conductor, it would possible to control the loss for a given field direction. In Fig 2 in [27] the effect of the aspect ratio on loss is shown for a rectangular conductor: the loss is basically proportional to the aspect ratio. This method is effective also in Bi-2223 tapes, i.e. adjusting the aspect ratio of the tape, from thin tape to square wire, according to the magnetic field direction, see for example [28] and the conclusion in [29].

In HTS solenoids, the largest contribution to the loss comes from the field component perpendicular to the tape at the coil ends. Therefore, if the tape was narrower and thicker (total current is unchanged) in these locations, the large radial loss contribution would be reduced, while the very small axial loss component would be increased. For example, a single tape could be replaced by a stack of three tapes (each one three times narrower), and the winding pack dimensions and operating current would be unchanged.

A non-twisted conductor is sometimes said to have enormous (infinite?) coupling losses, but this is not correct. The value of the critical twisting length $L_c = \sqrt{\dfrac{d \rho J_c}{dB/dt}}$, where $d$ is the superconductor thickness, $\rho$ the matrix resistivity, $dB/dt$ the magnetic field sweep rate, plays a major role in the determination of the loss regime. As discussed in [8] and [30], when the length of the sample (in non-twisted conductors), or the twist pitch (in twisted conductors) is $< L_c$, the coupling current is smaller than the screening current in the superconductor (the so-called uncoupled case). It follows that coupling losses are smaller than hysteretic losses. Coupling loss grows with the square of the twist pitch only till the twist pitch is the same order of magnitude of the $L_c$. When the sample length or the twist pitch is comparable or longer than $L_c$, the coupling current has the same magnitude than the screening current in the superconductor. The filaments are then fully coupled and the conductor behaves like a monolithic one with homogeneous current density. The coupling loss saturates to the hysteretic loss value of a monolithic conductor, the so-called saturated (coupling) loss. Saturated losses, like hysteretic losses, do not depend on frequency; the reason is that the time constant being extremely large, there is actually no dependence from the sweep rate. Analysis of the various loss regimes is reported also in [24].

Twisting is effective in reducing coupling losses only if the twist pitch is much smaller than $L_c$. Hysteretic losses can be reduced by choosing the finest possible filaments. In $Nb_3Sn$ and NbTi wires the filament diameter can be as small as few microns, while in coated conductor the effective width is always of the order of mm, three order of magnitude larger than in LTS. The losses in Bi-2223 tapes are the saturated losses of the monolithic conductor formed by the envelope of the filaments, and Bi-2223 magnets work fine even if the tape is operated in saturated loss regime. In Bi-2223 tapes with twisted filaments (and in $Bi_2Sr_2Ca_1Cu_2O_x$ wires), the losses are reduced only at low sweep rates (when twist pitch is $<L_c$), because the matrix resistivity is very low. Bi-2223 tapes with resistive barrier surrounding each filament were investigated in the 90's, but never reached industrial production.

In the non-twisted stack, the REBCO tapes are fully coupled, and the loss would be the same of a monolithic superconductor with rectangular cross section. Because of the very large aspect ratio, the rectangular stack can be treated as an infinite slab, and the calculation is similar to the one in [22] for a single tape. The loss (J/m$^3$) for an infinite slab (cycling between –B and +B) is:

$$Q = \begin{cases} \dfrac{2\mu_0}{3}\dfrac{B^3}{B_p} & B < B_p \\ \dfrac{2B_p}{\mu_0}\left(B - \dfrac{2}{3}B_p\right) & B > B_p \end{cases} \qquad (1)$$

In case of a single tape, $B_p = \mu_0 J_{sc}\dfrac{t}{2}$ for axial loss (field parallel to the tape), where $J_{sc} = \dfrac{I_c}{2wt}$, $t$=1 µm is the ceramic layer thickness and $w$=2 mm is the tape half width; the loss per unit length (J/m) is then $Q_z = Q2wt$. $B_p = \mu_0 J_e w$ for radial loss (field perpendicular to the tape), where $J_e = \dfrac{I_c}{2w2d}$ and $d$=0.15 mm is the total tape thickness including insulation; the loss per unit length (J/m) is then $Q_r = Q2wd$.

The field dependence of the tape critical current at 4.2 K is plotted in Fig. 1. In Fig. 2 the hysteretic loss (J/m) has been plotted as a function of field amplitude for a tape in axial configuration and for one in radial configuration. The instantaneous power loss is the time derivative of the curves in Fig. 2, divided by four (the loss for a ramp from 0 to $B$ is ¼ of the loss for AC field of amplitude $B$).

The calculation of saturated loss for the stack of three tapes is similar to the one used for calculating the hysteretic losses of one tape, using equation (1). The only modifications are the new dimensions and current densities of the stack. $B_p = \mu_0 J_e \dfrac{w}{2}$ for radial loss, where $J_e = \dfrac{I_c}{wD}$ and $w$=1.3 mm is the tape half width and $D$= 0.35 is the total stack thickness including insulation. $B_p = \mu_0 J_e \dfrac{d}{2}$ for axial loss, where $J_e = \dfrac{I_c}{wd}$, and $d$ =0.2 mm is the distance between the outermost ceramic layers.

The saturated loss (J/m) of the non-twisted stack is plotted in Fig. 2 together with the loss of the 4 mm wide tape. In long coils, where a large volume is exposed to axial field, the non-twisted stack may not provide the lowest loss, if used in the whole coil volume. Nevertheless, the region of large loss dissipation would be moved from the coil ends (where the critical current and margin is the lowest) to the central part. But in short coils, it may be advantageous (lower loss and lower magnetisation) to replace a wide tape with a stack of composed of few, narrow tapes. Application to graded coils will be discussed the next section.

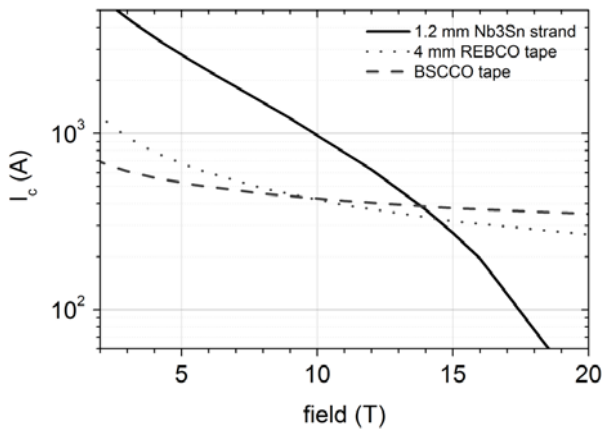

Fig. 1. Critical current versus field (at 4.2 K), for the wire and strands discussed in this paper.

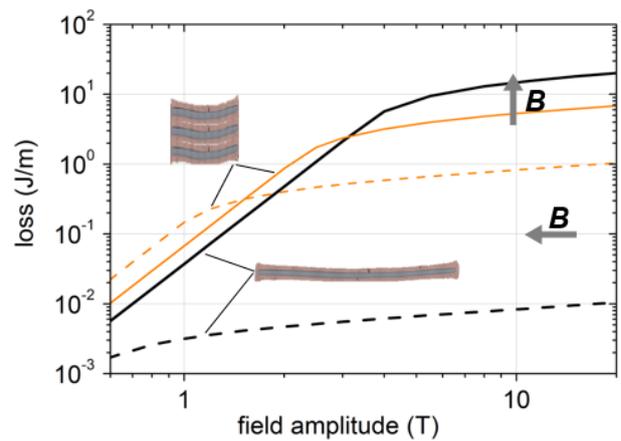

Fig. 2. Hysteretic loss (J/m) for a 4 mm wide tape and saturated loss for a stack of three 1.3 mm wide tapes. Both parallel (dashed lines) and perpendicular (solid lines) magnetic field orientation are shown.

A recent experiment [31] from the Shanghai University has demonstrated this concept. Two coils with similar dimensions were prepared: one was wound with a single 5 mm wide tape, the other with a stack of two 1.5 mm wide tape. The magnetization in the coil centre was found to be larger for the 5 mm tape coil than the one of the 2x1.5 mm tape coil. The goal of this experiment was not to reduce loss, but rather reduce the disturbance of the magnetization on the central field, which is probably a more important goal than loss reduction. A similar conductor, composed of two, non-twisted tape, was tested at CEA in a dipole. This will be discussed in section 4.3.

Only the Brookhaven Technology Group Inc. has dared to fabricate non-twisted stacks of more than two tapes [32]: their ExoCable™ is composed of a stack of exfoliated ceramic layers, each with the corresponding copper layer. A 10 m long cable was manufactured by assembling and soldering together a stack of eight exfoliated tapes (each 2.4 mm wide), resulting in a 1.2 mm thick conductor. The remnant magnetisation in the coil wound with such cable was lower than the ones of a coil wound with a 12 mm wide tape (see Fig. 14 in [32]). The measured hysteresis curves have been modelled considering the cable as a rectangular slab. This confirms that the ceramic layers are fully coupled and behave like a monolithic conductor, and that the method used to obtain Fig. 2 is valid. The rectangular slab model will be used to describe losses in thicker stacks in section 5.

For both the Shanghai University and Brookhaven Technology Group experiments, the magnetisation is proportional to the stack width, as expected, because the main contribution to the total magnetisation is the one from the screening current in the ceramic. For thicker stacks made of narrower tape, the effect of the field generated from saturated coupling current may not be negligible anymore.

### 4.3 Stack of non-twisted tapes for solenoids and dipoles

The discussion in the previous section and the experimental results of [31] and [32] suggest that the residual magnetization (and losses) in solenoids and dipoles could be reduced by decreasing the tape width at the coil ends, where the radial field component is larger. This would naturally lead to axial grading, and applications to solenoids and dipoles are briefly discussed.

In stand-alone coils wound with coated conductors, axial grading of the critical current (wider tape at the coils ends) is beneficial, because it allows to generate the same field with less material than in a non-graded coil (see for example [33]). The graded magnet in [33] was wound choosing 4.1 mm tapes in the mid-plane pancakes, and increasing the tape width up to 8.1 mm in the pancakes at the magnets ends. The disadvantages of using wide tapes at the magnet ends are large magnetization field and large loss. Both disadvantages can be eliminated by replacing the wide tape with stack of narrow tapes (for example 3 tapes each 3 mm wide), as schematically shown in Fig. 3 left.

The CEA block coil dipole [34] is also wound with a stack of two non-twisted tapes (each 12 mm wide). The coil is composed of a central double pancake and two smaller pancake on top and bottom, all wound with a two-in-hand conductor. The conductor is composed of two 12 mm wide tapes soldered to a central copper tapes and sandwiched between two CuBe tapes (see Fig. 3 in [34]). In this coil the screening currents perturbed the magnetic field in the dipole centre [35]. The magnetization could be reduced by replacing the top and bottom pancakes with pancakes wound with stack of few, narrow tapes, for example 3x8 mm and 6x4 mm. This configuration is schematically shown in Fig. 3 right.

Also regarding dipoles, Roebel cables have been considered for the construction of high field dipoles. Tapes in Roebel cables are uncoupled only when the field is parallel to the wide face of the tapes [36], instead, for perpendicular fields, the tapes are fully coupled. Therefore, a Roebel cable would have lower loss and lower magnetization than a non-twisted stack only in magnet sections where the field is mainly parallel to the tapes.

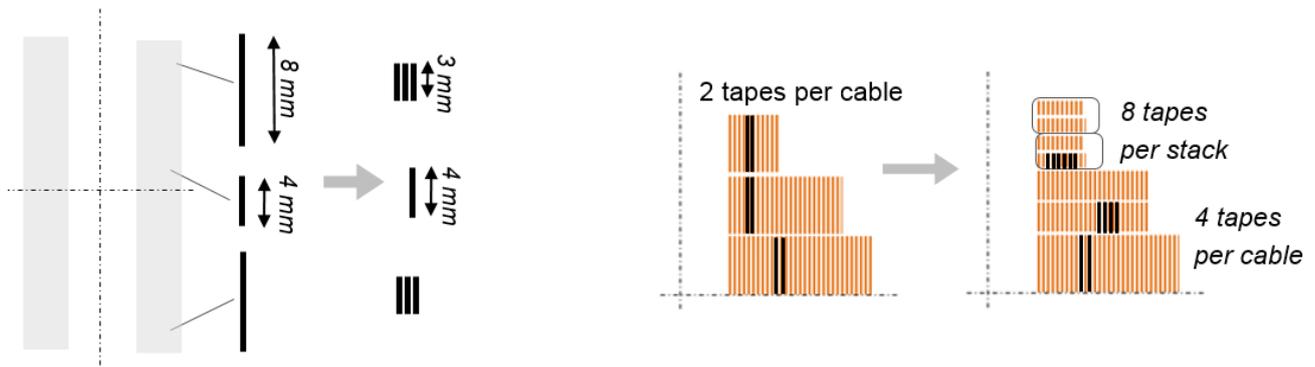

Fig. 3 Schematic illustration of the proposed replacement of wide tapes with non-twisted stacks of narrower tapes. Left: graded solenoid [Hahn]. Right: dipole [Durante].

### 4.4 BSCCO tapes and non-twisted stack of REBCO

The loss reduction strategy consisting in reducing the width of the tape in the direction perpendicular to the field is already exploited in Bi-2223 tapes. For example, Sumitomo produces "low loss" Bi-2223 tapes (non-twisted filaments), which are narrower than the standard Bi-2223 tapes. How would compare the losses of a non-twisted stack of three REBCO tapes (each 1.3 mm wide and 0.1 mm thick) with the one of a Bi-2223 tape? The losses have been calculated with the same procedure of the previous sections, using the corresponding dimensions and current density. When the loss per unit of critical current (J/mA) is considered (see Fig. 4), the non-twisted stack of REBCO tapes has comparable or lower loss than the Bi-2223 tape at any magnetic field (> 1 T) and for both orientations of the magnetic field. This suggests that a non-twisted stack of few REBCO tapes could be used in any magnet that was successfully operated with Bi-2223 tapes, i.e. dipoles, MRI coils and NMR inserts.

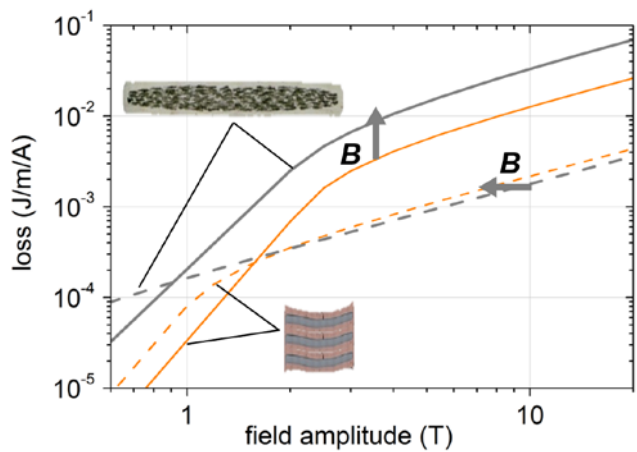

Fig. 4. Saturated loss divided by the critical current for a stack of tapes and for a BSCCO tape. Both parallel (dashed lines) and perpendicular (solid lines) magnetic field orientations are shown.

### 4.5 LTS tape magnets

What would happen if a coil like the HTS insert of Tohoku University had been built with LTS coated conductor? This may seem a strange question today, but $Nb_3Sn$ ribbons (12 mm wide, substrate <50 micron) have been available commercially from four different producers during the 60's and 70's, and several magnets were built with such ribbons. It has been reported [37] that those magnets often quenched at fields much lower than the one expected from the performance of short sample. The explanation was flux jumping, because LTS tapes are prone to flux jump instabilities when exposed to perpendicular fields of a Tesla or less. The solution was to add very thick stabilizing layer (for example very high purity aluminium), but better cooling was found to be more effective [38].

In section 4.2 it was shown that a solenoid wound with a tape has losses per unit volume which are order of magnitudes larger than the one wound with multifilamentary wire. This means that ramping losses of $Nb_3Sn$ ribbon magnets would be about two order of magnitude larger than in $Nb_3Sn$ multifilamentary magnets: probably the reported instabilities were not entirely due to flux jump, but also favoured from the large ramping loss.

## 5. Twisted and non-twisted cables for large magnets

Magnets for Fusion and for large detectors have operating currents exceeding 20 kA; this implies that hundreds of tapes must be assembled in a cable. Because of the very large number of tapes, the cable usually consists of sub-elements. The sub-elements considered so far are the stack of tape, the CORC and the Roebel cable; an overview of the present research activities is available in [3], sections 4 and 5. In the last few years, the most studied sub-element is the stack, which was first introduced by MIT [15].

### 5.1 Hysteretic and coupling loss in tape stacks and multifilamentary cables

As discussed in [24], the largest AC loss in a monolithic conductor (saturated loss) is simply proportional to the width of the conductor (in the plane perpendicular to the field direction). This value is found not only in monolithic conductors, but also in multifilamentary conductors when the filaments are fully coupled (saturated coupling loss, see section 4.2). Full coupling happens when the twist pitch is much longer than the critical twisting length or when the filaments are non-twisted. The effectiveness of the loss reduction strategy (for example twisting) is evaluated against the saturated loss.

We consider the effectiveness of twisting in two sub-cables: a twisted $Nb_3Sn$ sub-cable and a stack of coated conductors. They follow the design of two experimental conductors: the $Nb_3Sn$ sub-cable is employed in the conductor manufactured and tested by SPC for the EU-DEMO [14]; the square stack is the sub-cable of a flat conductor that is proposed also by SPC for the Central Solenoid of DEMO [39].

The $Nb_3Sn$ sub-cable is composed of 12 around 6 strands (1.2 mm $\varnothing$) cabled around a central copper wire. Cu:nonCu ratio is 1 and each strand is assumed to contain 2500 filaments of 10 µm diameter. The field dependence of the strand critical current is shown in Fig. 1. If neither the strands nor the filaments are twisted (coupled filaments), the saturated loss (J/m$^3$) can be calculated using the expression for a cylinder in transverse field (see [8], page 165-169), where the diameter is the one of the sub-cable and the current density is calculated over the whole cross section. If both strands and filaments are twisted (uncoupled filaments), the hysteretic loss is calculated with the same formula, but now the diameter is the filament diameter and the current density is the one in the filaments. In case of sufficiently low sweep rate, short twist pitch and high transverse resistance, the coupling losses are negligible, and the hysteretic loss is the only loss contribution. The losses for the two cases are plotted in Fig. 5: the benefit of twisting fine filaments and strands is an evident reduction of almost two order of magnitudes in losses at large field amplitude. In other words, filament hysteretic losses are much smaller than the saturated loss.

The stack of coated conductors is composed of 30 tapes, each 3.3 mm wide (tape $I_c$ in Fig. 1), resulting in a square stack; the field dependence of $I_c$ for one tape is shown in Fig. 1. The saturated loss (J/m$^3$) for the non-twisted stack can be calculated from $\chi''$, the imaginary part of the complex susceptibility, as described in [40]: $Q = \pi\mu_0 H^2 \frac{\chi''}{\chi_0}\chi_0$, where $\chi_0$ is equation (13) in [Pardo]. The $\chi''/\chi_0$ values as a function of $h = H/H_p$ can be found in table 2 in [40] for rectangular conductors of various aspect ratio.

$H_p = \frac{J_e b}{\pi}\left[\frac{2a}{b}\arctan\frac{b}{a} + \ln\left(1 + \frac{a^2}{b^2}\right)\right]$ is the penetration field for a rectangular conductor, where $a$ is the stack half width and $b$ is the stack half height, here $a=b=1.65$ mm; $J_e = \frac{I_c}{2a2b}$. The saturated loss (J/m) is obtained by dividing $Q$ by the stack cross section and is plotted in Fig. 5 (dashed green line).

If the stack is twisted, the loss is only $2/\pi \sim 0.64$ of the loss of the non-twisted stack, as demonstrated in [15], section 7.2.2; a similar analysis has been carried out in [41] and in [42]. The reason is that a twisted stack has negligible hysteretic loss only where the tapes are almost parallel to the magnetic field, which occurs only twice along a pitch length (see fig. 1 in [42]); along most of the length, twisted and non-twisted stacks have similar losses. In first approximation, the effective width of a twisted tape is about 0.64 times the width of the tape; for a more precise loss evaluation one should take into account also the angular dependence of the critical current. In other words, the hysteretic loss of a twisted stack is almost as large as the saturated loss of a non-twisted stack.

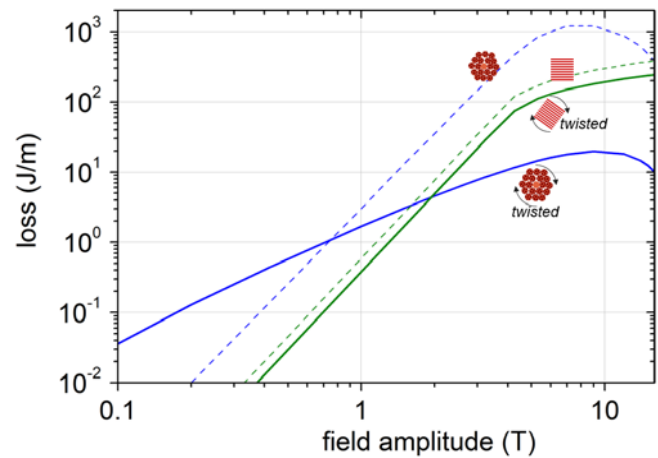

Fig. 5. Hysteretic loss for the stack of tapes and the Nb$_3$Sn sub-cable. The twisted (solid line) and non-twisted (dashed line) cases are shown.

Large hysteretic losses in stack of tapes have been confirmed experimentally. The magnetization loop of a non-twisted stack (28 tapes, 3 mm wide) was measured at Frascati (see Fig. 9 in [43]): from the area of the magnetization loop (from -12 T to 12 T) the hysteretic loss can be estimated to be about 260 J/m, which is in agreement with Fig. 5. The corresponding energy loss density would be about 30 MJ/m$^3$. The magnetization loop was also measured at an angle of 45° between the stack and the magnetic field (see Fig.11 in [43]). The loop area, and thus the loss, is just slightly lower than the one for the field orientation perpendicular to the tapes, proving that hysteretic losses in twisted stacks are only marginally lower than in non-twisted ones.

To summarise, twisting is very effective (few orders of magnitude) in reducing losses for the LTS sub-cables, because when twisting (i.e. magnetically decoupling) round, fine filaments, the effective diameter drops from few millimetres to few microns (two or three order of magnitudes smaller). Instead, twisting a stack of tapes is a very ineffective loss reduction strategy (less than 36% reduction), because the effective width decreases only marginally. If it was possible to subdivide the tape into very narrow (few tens of micron) filaments, the so-called striation process, the loss of a twisted stack could be drastically lower than the one of a non-twisted stack (see for example [27]).

## 5.2 AC loss in large tape stack cables

Before evaluating the loss in cables, field sweep amplitudes (ΔB) and sweep rates should be discussed. The values here and in the following sections are from DEMO EUROfusion (2015 version), are only indicative and to be used as example for loss assessment.

In large tokamaks, the Toroidal Field (TF) magnets will be ramped at sweep rates between 0.001 and 0.01 T/s and field sweep (ΔB) from 0 T to 12 T (up to 20 T for compact, HTS tokamak). This range of sweep rate is the same of small laboratory magnet, NMR, and even the LHC dipoles. Instead, the Central Solenoid (CS) operation is complex, see for example [44]. The sweep field is usually from about -12 T to 12 T, but with the use of HTS, fields up to 17 T have been considered. Assuming a 16 T CS field, as in [40], the three most demanding CS operation modes, regarding AC losses, are:
- Dwell phase: field sweep from about -16 T to about 16 T; sweep rate between 0.05 and 0.1 T/s.
- Breakdown phase: ΔB is about 1 T at peak field (16 T); sweep rate is about 1 or 1.5 T/s.
- Plasma current ramp-up (PCRU) phase: field sweep from 16 T to about 0 T; sweep rate between 0.2 and 0.3 T/s.

Large fusion conductors are tested over a broad range of sweep rates but at small AC field amplitude (<0.5 T), for example at Twente University or in SULTAN at SPC. In the SULTAN facility the static background field is up to 10 T, with a maximum AC field amplitude of 0.4 T; the sweep rate can vary from about 0.1 T/s to about 3 T/s, roughly corresponding to frequencies between 0.1 and 3 Hz.

Losses on an HTS experimental cable composed of twisted stacks have been measured in SULTAN [45], following the same procedure and in the same conditions of ITER conductors. The HTS conductor was composed of rectangular stacks, whose hysteretic losses are similar to the ones of the square stack in Fig. 5. These measurements showed that hysteretic losses were negligible and that the coupling losses were moderately larger than in $Nb_3Sn$ CICC. This result is not in contrast with Fig. 5, as explained below.

The total energy deposited by coupling loss during a field ramp is proportional to the sweep rate. Therefore, the value measured in the SULTAN test can be directly used for assessing coupling losses during operating phases, if measured at the corresponding sweep rate. Instead, the energy deposited by hysteretic loss is proportional to the AC field amplitude $B$, and more precisely to $B^3$ when the field is lower than the penetration field and to $B$ when the field is larger than the penetration field. In twisted multifilamentary cables the penetration field is <0.1 T; therefore the extrapolation of hysteretic losses from the SULTAN test (0.4 T field amplitude), to the operation of the CS (> 16 T) is linear (see Fig. 5). In a stack of tapes (twisted or not), the penetration field is about 2 T to 3 T (see for example Fig. 9 in [43]); therefore the extrapolation follows the third power of the field up to $B_p$, and then is linear (see Fig. 5). It follows that for low field amplitudes (<0.5 T), the stack has the lowest hysteretic losses but they grow much faster with field, and at field >10 T, the HTS twisted stacked cable has hysteretic losses ten times larger than the LTS counter-part; this result was already reported in [Takayasu]. In addition, one should not forget that twisting the stack reduce hysteretic losses only by 30%, but it also introduce coupling losses among the tapes in the stack.

From the magnetization curve of Fig. 9 in [43], the loss can be estimated at about 260 J/m for one (non-twisted) stack, for an AC field of 12 T. This is in agreement with Fig. 5, where the hysteretic loss for a twisted stack is about 200 J/m for 12 T field amplitude and 230 J/m for 16 T field amplitude. This last value should be multiplied by 12 (number of stacks) to obtain the hysteretic loss of a full cable (assuming there is no magnetic interaction between the stacks). During the PCRU phase, the total energy deposited by hysteretic loss is thus 200 J/m x 12, divided by 4 (ramping from 16 T to 0 T instead of 16 T AC amplitude), which corresponds to about 600 J/m. The average power loss of 7.5 W/m is then obtained by dividing 600 J/m by the duration of the PCRU phase (80"). The energy deposited during the dwell phase is about 1200 J/m and the corresponding average power loss (over 600" of duration) is about 2 W/m. These values are much larger than the power coupling loss considered in [44] for the HTS cable: 1.9 W/m for the PCRU phase and 0.1 W/m for the dwell phase. One more important difference between coupling and hysteretic power losses is that hysteretic losses are not constant with time, like the coupling ones, but grows with the magnetic field, as it is the case for magnets build with single tape (see for example Fig. 6 in [22]). The values of 7.5 W/m and 2 W/m are only average values; actually the peak hysteretic power loss can been estimated (from Fig. 5) to be >20 W/m for the PRCU phase and >10 W/m during the dwell phase, that is one to two order of magnitudes larger than the coupling power losses.

The coupling and hysteretic losses for the four operating cases are reported in Table 2. The coupling loss is calculated as in [44], while the hysteretic losses are derived from Fig. 5. The hysteretic loss during breakdown is calculated with 1 T AC field amplitude but using the current density at 15 T instead of 1 T. The losses are also plotted in Fig. 6, which is a kind of "loss map" similar to Fig. 10 in [24]. In Fig. 6 the grey shaded areas indicate sweep rate and field sweep amplitude for the four operating modes of CS magnet; for each mode the coupling and hysteretic losses are indicated. In Fig. 7 the power losses are plotted; here the power coupling losses are proportional to the square of the sweep rate.

*Table 2. Loss estimation for HTS flat cable (12 twisted stack, see [Bykowsky2018]) in Central Solenoid.*

|  | Dwell | Breakdown | PCRU | burn |
|---|---|---|---|---|
| ΔB | -16 T to 16 T | 16 T to 15 T | 15 T to -2 T | -2 T to -16 T |
| Sweep rate | 0.2–0.3 T/s | 1–2 T/s | 0.04–0.06 T/s | 0.002 T/s |
| duration | 600" | 1" | 80" | 7200" |
| Hysteretic loss | 1200 J/m | 35 J/m | 610 J/m | 600 J/m |
| Coupling loss | 70 J/m | 90 J/m | 150 J/m | 1.2 J/m |
| Average hysteretic power loss | 2 W/m | 35 W/m | 7 W/m | 0.1 W/m |
| Coupling power loss | 0.1 W/m | 95 W/m | 2 W/m | $10^{-4}$ W/m |

In [2], it is claimed that coupling loss is the main loss contribution in HTS stack cables; this statement is inaccurate, because its validity is limited to field amplitudes smaller than the penetration field of the stack. At such low fields (< 0.4 T) most of the stack is screened from the magnetic field, and the losses are the ones of few tapes located on the stack exterior. Instead, when the ΔB exceeds few Tesla (full penetration), the main source of loss in tape stack cables will be the hysteretic one. Therefore the thermo-hydraulic analysis in [Dembrowska] underestimates the temperature increment, because it takes into account only coupling loss and neglects the contribution from the hysteretic loss, which is the largest one during the plasma current ramp-up and dwell phases (large ΔB, see Table 2). Ramping losses in large HTS fusion cables are much higher than in LTS ones, in the same way as ramping losses in small HTS solenoids are much higher than in LTS solenoids (see section 4.1).

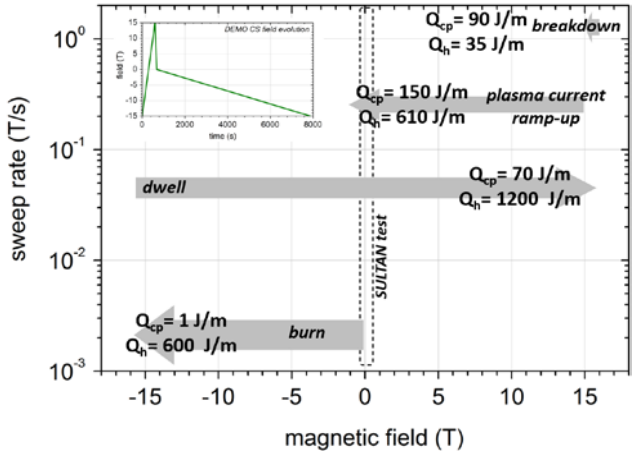 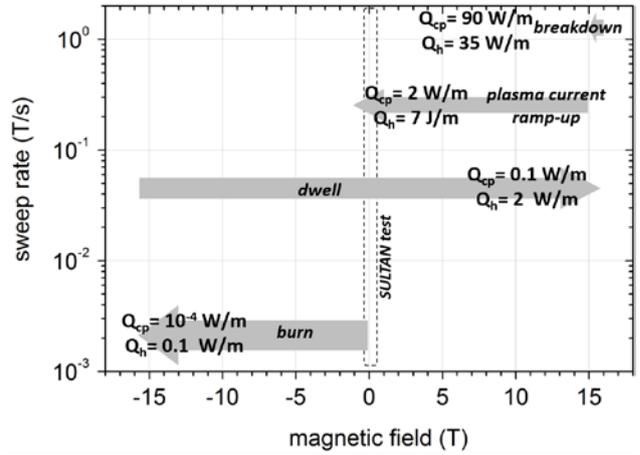

Fig. 6. Loss map for a twisted stack cable. Grey regions indicates the different sweep rates and field variations for the different CS operation phases. Coupling loss ($Q_{cp}$) and hysteretic loss ($Q_h$) are indicated. The inset shows the temporal evolution of the magnetic field in the DEMO CS.

Fig. 7. Power Loss map for a twisted stack cable. Grey regions indicates the different sweep rates and field variations for the different CS operation phases. Coupling power loss ($P_{cp}$) and average power hysteretic loss ($P_h$) are indicated.

### 5.3 Twisted and non-twisted conductors for fusion magnets

As mentioned in section 4.2, the aspect ratio of the conductor (here a non-twisted stack of tape) can be selected to lower the loss, depending on the main field direction in the magnet, or in a certain magnet section. In this section, this strategy will be applied to conductors for CS and TF magnets for large tokamaks.

#### *5.3.1 Conductors for CS*

The EU-DEMO (2015 version) CS is composed of five solenoid modules (from top to bottom: CSU3, CSU2, CS1, CSL2 and CSL3). The modules are powered independently, but the field is raised and decreased in all modules almost at the same time (see [Wesche2018]). It follows that, at high magnetic field, the field direction in the three central modules is almost parallel to the solenoid axis (<5° misalignement). Therefore, for any magnetic field, the critical current of a tape oriented parallel to the solenoid axis is at least two times higher than the one of a twisted tape.

One option considered at SPC for the central solenoid is a flat HTS cable composed of twisted square stacks (3.3 mm x 3.3 mm), as shown in Fig. 8 (see also Fig. 7 in [39]). In the three central modules (CS1, CSU2, CSL2) the number of tapes can be reduced by 50%, because the critical current is at least twice than for the twisted stack, as discussed above. Two possible ways to arrange the tapes, maintaining the parallelism with the coil axis, are the following. 1) Twelve rectangular stacks (aspect ratio is two) arranged in a Robel bar, each stack containing half the number of tapes than the twisted stack cable, as shown in Fig. 8. 2) Assembling all the tapes in a single monolithic non-twisted stack, 26 mm wide and 2.6 mm thick, see Fig. 8.

The hysteretic loss of the flat cable and the saturated loss of the Roebel can be estimated by multiplying by 12 the respective stack loss. This procedure does not take into account the magnetic interaction between neighbouring strands, which is important at field lower than the penetration field of one stack; effects of the interactions are discussed in [47]. The flat cable and the Roebel cable would have the same twist pitch; therefore the coupling losses are also similar, but are not considered here. In the monolithic stack, the only loss contribution is the saturated loss and it is calculated using the susceptibility for rectangular conductors [40].

The losses due to radial fields (<1 T), i.e. perpendicular to the wide face of the conductors, are several orders of magnitude smaller than the ones due to the axial field (parallel to the wide face), and can be neglected for all conductors. The axial field losses are shown in Fig. 9. The non-twisted stack Roebel has lower losses than the twisted stack cable at any field amplitude; the loss of the monolithic stack are comparable to the one of the twisted stack cable at large field amplitude. Therefore, non-twisted stack conductors similar to these ones should not pose any issue from point of view of losses in the central modules of a CS magnet.

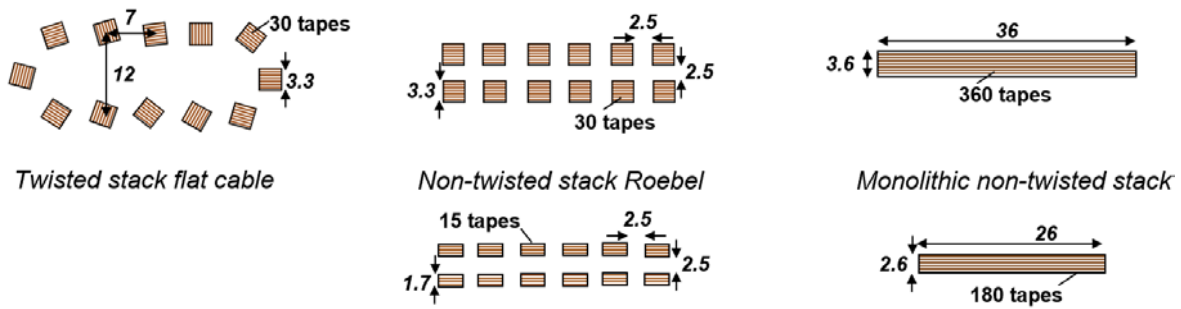

Fig. 8. Sketches of the three types of cables: twisted stack flat cable, non-twisted stack Roebel and (non-twisted) monolithic stack. Copper for quench protection, space for coolant and jacket are omitted for clarity. Stack separations are not used for AC loss estimation.

In the top and bottom modules (SCU3 and CSL3), the number of tape can not be reduced because the radial component is > 5 T. Therefore, the Roebel cable and the monolithic stack should now contain the same number of tapes than the twisted stack flat cable, as shown in Fig. 8. The saturated losses are now calculated with the new stack dimensions and engineering current densities and are plotted in Fig. 10 for both axial and radial field components; the maximum field ranges are indicated by the grey regions. The non-twisted stack Roebel has 30% higher hysteretic losses than the twisted stack cable in both orientations. Eventually this extra power may be removed with the same electrical power if the operating temperature is increased. This disadvantage should be weighed against the advantages, for example much higher tolerance against transverse pressure. The monolithic stack has clearly much higher hysteretic losses than the other cables, because the loss due to the radial field is about 10 times larger. Of course, losses can be reduced by reducing the aspect ratio, but the minimum bending strain would increase.

*5.3.2 Conductors for TF*
In EUROfusion DEMO (2015 version), the peak field on the TF winding pack should be <14 T, within reach of $Nb_3Sn$ conductors. Nevertheless, HTS cables are also investigated; one motivation is that, if the price of coated conductors is sufficiently reduced (about an order of magnitude), it would be cheaper than $Nb_3Sn$ even at low fields. The TF magnet can be slowly charged (sweep rate <0.01 T/s), but it will be subjected to the stray field of the CS and PF magnet. Nevertheless, the CS stray field component normal to the TF conductor is about 300 times smaller than the central field of the CS; therefore the sweep rate during breakdown (1 T/s in CS centre) will be <0.005 T/s (comparable to the charging field rate), and the related loss can be neglected. The PF stray field could be as high as 1 T, but the sweep rate is not yet known.
In the case of the TF magnet, the alignment of the tapes with the magnetic field is difficult to obtain, because the magnetic field is parallel (<5°) to the winding pack only in less than half of winding pack cross section. Therefore, the number of tapes should be the same as in the twisted stack flat cable.

For the estimation of the loss, the data plotted in Fig. 10 should be used, the dark grey regions indicating the field of interest have been adapted to the TF. As it was the case for the CS top and bottom modules, the non-twisted stack Roebel has always 30% higher hysteretic losses than in the one of the twisted stack flat cable. The loss of the monolithic stack can be 5 times higher than in the twisted stack flat cable in some locations of the winding pack; when the losses over the whole winding pack are summed up, the total loss is two times higher than the one of the twisted stack flat cable.

During TF magnet charging, there is no nuclear heat load, therefore the extra cooling capacity could be instead used to remove the larger hysteretic losses. Thermo-hydraulic analysis should be carried out to quantify the disadvantage of large losses (from 30% to 2 times higher). The main potential advantage of non-twisted configurations is that they will likely have better tolerance against transverse pressure. In the three central modules of the CS solenoid, a further, but minor advantage is the reduction in the number of superconducting material and thus in cost, because even at present price (20 €/m for a 4 mm wide tape), the total price of the tape for the twisted stack cable would be less than 40 million €. Saving 10–15 million with the non-twisted stack is not much compared to the >10 billion € for the whole DEMO power plant.

In high field compact tokamaks, considered by Tokamak Solutions and Commonwealth Fusion, the peak field would exceed 20 T at 20 K, and the whole winding pack contains only coated conductors. The high operating temperature mitigates the disadvantage of non-twisted designs (large loss). At the same time, the advantages (higher current density and higher strength against transverse load) gain importance because of the high operating magnetic field and temperature. Therefore, the use of non-twisted stacks in these tokamaks could have much bigger impact than in the large, low field magnets of EUROfusion DEMO.

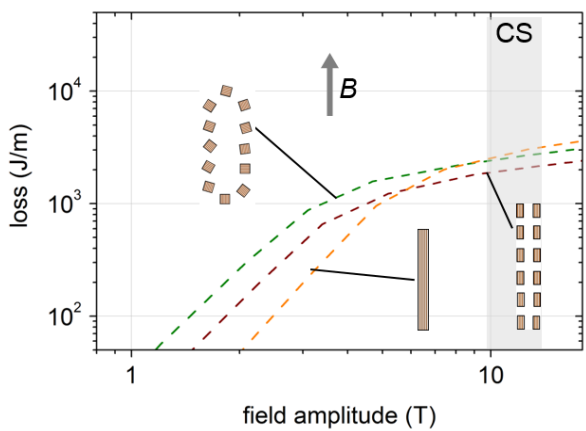

Fig. 9. Hysteretic and saturated losses in the central CS modules for the three cables. Magnetic field parallel to the wide face of the cable (axial field losses). The grey band indicate the maximum range of field in CS magnets.

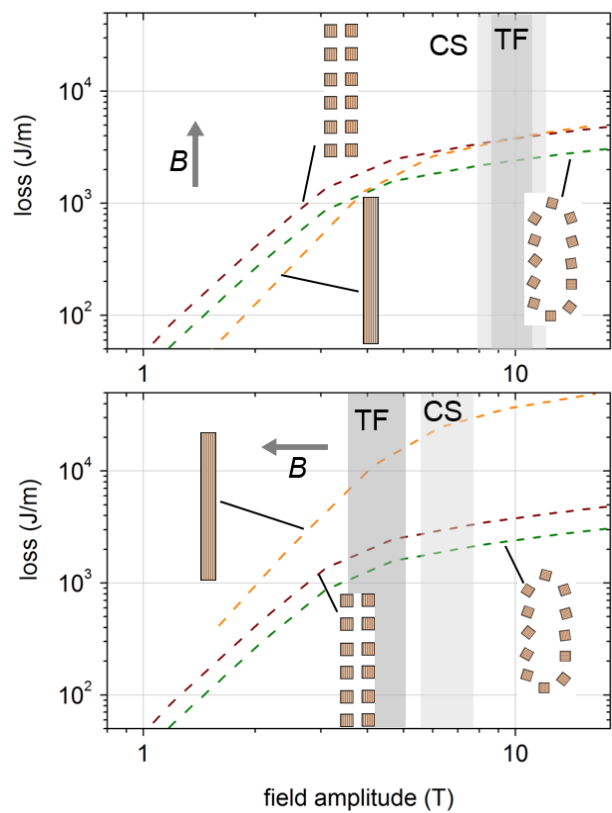

Fig. 10. Hysteretic and saturated losses in the outermost CS modules for the three cables. Top: Magnetic field parallel to the wide face of the cable (axial field losses). Bottom: magnetic field perpendicular to the wide face of the cable

(radial field losses). The grey band indicate the maximum range of field in CS (light grey) and TF (dark grey) magnets.

## 6. Conclusions

Numerical calculations have shown that twisting a stack of tapes has negligible effect on the inductance variation among the tapes (few % for a large coil); the mismatch in inductance originates from mutual inductances. Even twisted filaments in LTS strands do not have the same inductance. This indicates that in relatively compact conductors (few mm in cross section size), transposition (zero inductance variation among the elements) and even twisting are not required.

High field magnets are usually built with a single HTS tape. These magnets have intrinsically higher losses than the ones built with fine multifilamentary wires, but they can be operated even if the instantaneous power losses are high, because of the large stability margin. If the tape is replaced by a stack of few, narrower tapes (for example 3 tapes 3 mm wide instead of a single 12 mm wide tape) at the coil ends, the losses and the field disturbance from screening currents in the superconducting layer can be reduced. This is valid not only for solenoids but also for dipoles. A stack of three, narrow (3 mm) REBCO tapes have losses comparable to BSCCO tapes, and can therefore be used to wind any kind of magnets that has been wound with BSCCO tapes.

LTS strands must be composed of fine twisted filaments because of low stability. Instead, HTS do not suffer from any type of instability. In twisted cable composed of fine multifilamentary wires, the hysteretic loss is several order of magnitudes lower than the saturated loss of a non-twisted one. Instead, the hysteretic loss of a twisted stack of tapes is only marginally lower (-36%) than the saturated loss of a non-twisted stack. Therefore twisting a tape stack is a very inefficient loss reduction strategy. Rather than twisting, modifying the stack aspect ratio of non-twisted stack can reduce losses. These considerations suggest that stacks of HTS tapes do not need to be twisted, because stability, inductance and AC losses are not much affected by twisting. This is not a new result in contrast with the classical LTS designs, but is a consequence of critical state model and stability analysis (over 60 years old) applied to HTS tape stacks.

Non-twisted stack of tapes could be used to wind a variety of magnets, from small, high field solenoids and dipoles (stack composed of 2 to 5 tapes) to large magnets for detectors and fusion (hundreds of tapes). Additional potential advantages of non-twisted stack designs are higher tolerance to transverse stresses and increase in critical current density Specific aspects of non-twisted designs should be further investigated, for example manufacturing and thermo-hydraulic behaviour.

In general, copying LTS magnet features (for example impregnation, cable compaction, temperature margin, cooling method, protection) in HTS magnets may lead to ineffective solutions, because HTS are radically different from LTS: much larger temperature margin, highly anisotropic transport and mechanical properties. HTS magnets should be designed considering the specific features of HTS and design choices must always be motivated by physical analysis, not simply copied from past LTS cables and conductors.